\documentclass[sn-mathphys]{sn-jnl}

\usepackage{graphicx}%
\usepackage{multirow}%
\usepackage{amsmath,amssymb,amsfonts}%
\usepackage{amsthm}%
\usepackage{mathrsfs}%
\usepackage[title]{appendix}%
\usepackage{xcolor}%
\usepackage{textcomp}%
\usepackage{manyfoot}%
\usepackage{booktabs}%
\usepackage{algorithm}%
\usepackage{algorithmicx}%
\usepackage{algpseudocode}%
\usepackage{listings}%
\usepackage{gensymb}
\usepackage{overpic}
\usepackage{ulem}
\usepackage{enumerate}
\usepackage{makecell}
\usepackage{lipsum}
\usepackage{siunitx}
\usepackage{lineno}
\usepackage{upgreek}
\usepackage{color}
\usepackage[figuresright]{rotating}

\theoremstyle{thmstyleone}%

\theoremstyle{thmstyletwo}%

\theoremstyle{thmstylethree}%

\begin{document}

\title[ ]{Pitch Angle Measurement Method based on Detector Counts Distribution. -I. Basic conception}

\author[1,2]{\fnm{Chenwei} \sur{Wang}}
\author*[1]{\fnm{Shaolin} \sur{Xiong}}\email{xiongsl@ihep.ac.cn}
\author[4,5]{\fnm{Hongbo} \sur{Xue}}
\author[4,5]{\fnm{Yiteng} \sur{Zhang}}
\author[4,5]{\fnm{Shanzhi} \sur{Ye}}
\author[6]{\fnm{Wei} \sur{Xu}}
\author[1,2]{\fnm{Jinpeng} \sur{Zhang}}
\author[1]{\fnm{Zhenghua} \sur{An}}
\author[7]{\fnm{Ce} \sur{Cai}}
\author[1,2]{\fnm{Peiyi} \sur{Feng}}
\author[1]{\fnm{Ke} \sur{Gong}}
\author[1,8]{\fnm{Haoxuan} \sur{Guo}}
\author[1]{\fnm{Yue} \sur{Huang}}
\author[1]{\fnm{Xinqiao} \sur{Li}}
\author[1,2]{\fnm{Jiacong} \sur{Liu}}
\author[1]{\fnm{Xiaojing} \sur{Liu}}
\author[1]{\fnm{Xiang} \sur{Ma}}
\author[1]{\fnm{Liming} \sur{Song}}
\author[1,2]{\fnm{Wenjun} \sur{Tan}}
\author[1]{\fnm{Jin} \sur{Wang}}
\author[1]{\fnm{Ping} \sur{Wang}}
\author[1,2]{\fnm{Yue} \sur{Wang}}
\author[1]{\fnm{Xiangyang} \sur{Wen}}
\author[9]{\fnm{Shuo} \sur{Xiao}}
\author[1,10]{\fnm{Shenglun} \sur{Xie}}
\author[1]{\fnm{Yanbing} \sur{Xu}}
\author[1,2]{\fnm{Wangchen} \sur{Xue}}
\author[1]{\fnm{Sheng} \sur{Yang}}
\author[1,2]{\fnm{Zhenghang} \sur{Yu}}
\author[1]{\fnm{Dali} \sur{Zhang}}
\author[1,11]{\fnm{Wenlong} \sur{Zhang}}
\author[1,12]{\fnm{Peng} \sur{Zhang}}
\author[1,2]{\fnm{Shuangnan} \sur{Zhang}}
\author[1,2]{\fnm{Yanqiu} \sur{Zhang}}
\author[1,2]{\fnm{Yanting} \sur{Zhang}}
\author[1]{\fnm{Zhen} \sur{Zhang}}
\author[1]{\fnm{Xiaoyun} \sur{Zhao}}
\author[1,2]{\fnm{Chao} \sur{Zheng}}
\author[1]{\fnm{Shijie} \sur{Zheng}}

\affil*[1]{\orgdiv{Key Laboratory of Particle Astrophysics}, \orgname{Institute of High Energy Physics, Chinese Academy of Sciences}, \orgaddress{ \city{Beijing}, \postcode{100049}, \country{China}}}
\affil[2]{\orgdiv{University of Chinese Academy of Sciences}, \orgname{Chinese Academy of Sciences}, \orgaddress{ \city{Beijing}, \postcode{100049}, \country{China}}}
\affil[3]{\orgdiv{School of Computer and Information}, \orgname{Dezhou University}, \orgaddress{ \city{Dezhou, Shandong}, \postcode{253023}, \country{China}}}
\affil[4]{\orgdiv{State Key Laboratory of Space Weather}, \orgname{National Space Science Center, Chinese Academy of Sciences}, \orgaddress{ \city{Beijing}, \postcode{100190}, \country{China}}}
\affil[5]{\orgdiv{Key Laboratory of Solar Activity and Space Weather}, \orgname{National Space Science Center, Chinese Academy of Sciences}, \orgaddress{ \city{Beijing}, \postcode{100190}, \country{China}}}
\affil[6]{\orgdiv{Electronic Information School}, \orgname{Wuhan University}, \orgaddress{ \city{Wuhan, Hubei}, \postcode{430072}, \country{China}}}
\affil[7]{\orgdiv{College of Physics}, \orgname{Hebei Normal University}, \orgaddress{ \city{Shijiazhuang, Hebei}, \postcode{050024}, \country{China}}}
\affil[8]{\orgdiv{Department of Nuclear Science and Technology}, \orgname{School of Energy and Power Engineering, Xi’an Jiaotong University}, \orgaddress{ \city{Xi’an}, \postcode{710049}, \country{China}}}
\affil[9]{\orgdiv{School of Physics and Electronic Science}, \orgname{Guizhou Normal University}, \orgaddress{ \city{Guiyang}, \postcode{550001}, \country{China}}}
\affil[10]{\orgdiv{Institute of Astrophysics}, \orgname{Central China Normal University}, \orgaddress{ \city{Wuhan, HuBei}, \postcode{430079}, \country{China}}}
\affil[11]{\orgdiv{School of Physics and Physical Engineering}, \orgname{Qufu Normal University}, \orgaddress{ \city{Qufu, Shandong}, \postcode{273165}, \country{China}}}
\affil[12]{\orgdiv{College of Electronic and Information Engineering}, \orgname{Tongji University}, \orgaddress{ \city{Shanghai}, \postcode{201804}, \country{China}}}

\abstract{
As an X-ray and gamma-ray all-sky monitor aiming for high energy astrophysical transients, Gravitational-wave high-energy Electromagnetic Counterpart All-sky Monitor (GECAM) has also made a series of observational discoveries on burst events of gamma-rays and particles in the low Earth orbit. 
Pitch angle is one of the key parameters of charged particles traveling around geomagnetic field. However, the usage of the GECAM-style instruments to measure the pitch angle of charged particles is still lacking.
Here we propose a novel method for GECAM and similar instruments to measure the pitch angle of charged particles based on detector counts distribution. The basic conception of this method and simulation studies are described. With this method, the pitch angle of a peculiar electron precipitation event detected by GECAM-C is derived to be about 90$^\circ$, demonstrating the feasibility of our method. 
We note that the application of this method on GECAM-style instruments may open a new window for studying space particle events, such as Terrestrial Electron Beams (TEBs) and Lightning-induced Electron Precipitations (LEPs).
}


\maketitle

\section{Introduction} \label{sec:intr}

Space charged particle events are widely observed in space by instruments on satellites, including various particle precipitation events \cite{LEP_SPA_Voss1984, NWC_Zhang_2016, GRL_Shumko2021,NC_zhang_2022}, Terrestrial Electron Beams (TEBs) \cite{GBM_TEB,GECAM_TEB,GBM_TGF_with_TEB,TEB_ASIM}, and others. 
Charged particle precipitation is often observed during geomagnetic storms and substorms, and plays an important role in the energy transfer of the Earth’s magnetosphere \cite{Osc_Bortnik2008}. 
The precipitating charged particles, with energy ranging from eV to MeV, significantly affect ionospheric chemistry and propagation of electromagnetic waves, often leading to interference or disruptions in communication systems and serious threat to the safety of spacecraft. 
The observation of charged particle precipitation events is essential to understand their generation and propagation processes, which will help to prevent their impact on the development and utilization of space.

Energetic charged particles can be accelerated and scattered by different wave‐particle interactions that lead to precipitation. Some wave modes are expected to produce specific patterns of pitch angle distributions, such as magnetospheric waves \cite{Ni_PA_2020}. 
As an essential form of energy transport in the magnetosphere, magnetic reconnection is also an efficient process to produce the suprathermal electron population, and the evolving pitch angle distribution can indicate the specific mechanism of reconnection \cite{Hoshino_reconnection_2001,Wang_reconnection_2010}. 
The simulation of TGF and TEB production shows that the duration of TEB is determined by the wide range of pitch angle of electrons, and predicts that the pitch angle distribution of electrons evolves as a function of time, and electrons with smaller pitch angles arrive earlier while electrons with larger pitch angle arrive later \cite{Dwyer_TEB_2008,Carlson_TEB_2009}. 
Therefore, pitch angle is one of the most important parameters of charged particles. The measurement of pitch angle can provide critical information for the generation, acceleration, and propagation of these phenomena. 

The measurement of pitch angle can be divided into two processes: one is measuring the angular distribution of the population of charged particles, and the other is measuring the magnetic field. 
Typically, the measurement of the angular distribution of particles is accomplished by restricting a narrow field of view using structures such as collimators. This can be achieved either by utilizing multiple detectors oriented in different directions to measure the angular distribution \cite{HEPP} or by using a narrow FoV detector to scan directions \cite{ELFIN_2020}, obtaining the flux at different angles. Recently, Reid et al. have proposed another new conception of pitch angle measurement by using coded aperture imaging \cite{pith_angle_overview}. 

In the history of charged particle detection, apart from the specialized space physics instruments such as the Van Allen Probes \cite{VAP_2013}, high-energy astronomical telescopes have also made significant contributions, particularly wide-field gamma-ray monitors such as the Burst and Transient Spectrometer Experiment (BATSE) onboard Compton Gamma Ray Observatory (CGRO) \cite{INS_CGRO_1994}, Gamma-ray Burst Monitor (GBM) onboard \textit{Fermi} \cite{INS_GBM_Meegan2009} and GECAM \cite{GECAM_TEB}. Thanks to the high detection efficiency brought by the design of a large field of view, these instruments aiming for time-domain astronomy have detected many special charged particle events \cite{GBM_TGF_with_TEB,GECAM_TEB}. Sarria et al. have tried to obtain the pitch angle of TEB from the temporal information indirectly by \textit{Fermi}/GBM data \cite{GBM_pitch_by_time}. However, due to the lack of collimating structures, people have long overlooked their inherent ability to measure the pitch angle.

Hence, we proposed a new pitch angle measurement method based on counts distribution (PACD) for GECAM, which is inspired by the counts distribution localization method for astrophysical bursts \cite{YIZ_LOC_GEC_2023}, and will expound the extensions, optimizations, and applications of PACD method in a series of articles. 
As the first paper of this series, the main purpose of this paper is to explain the basic conception of the PACD method, and this paper is organized as follows. We present the basic information of GECAM and the detailed description of the primary PACD method in Section\,\ref{sec:ins}, with an ideal assumption. Then we performed the method on a charged particles event (labeled as tn240719\_200130) detected by GECAM-C and the pitch angle measurement result is shown Section\,\ref{sec:obs}. Discussion and summary are given in Section\,\ref{sec:sum}.

\section{Instrument and Method} \label{sec:ins}
GECAM is a constellation dedicated to the monitor all-sky high energy astrophysical transients as well as terrestrial transients. At the time of writing, GECAM is composed of four instruments, GECAM-A, GECAM-B \cite{GEC_INS_Li2022}, GECAM-C \cite{HEBS_INS_Zhang2023} and GECAM-D \cite{GTM_INS_wang2024}. As a versatile instrument, GECAM has made a series of discoveries in high energy transient phenomena, including Gamma-Ray Bursts (GRBs) \cite{GECAM_GRB_1,GECAM_GRB_2,GECAM_GRB_3}, Soft Gamma-ray Repeaters (SGRs) \cite{GECAM_SGR_1,GECAM_SGR_2}, high energy counterpart of Gravitational Wave (GW) and Fast Radio Burst (FRB) \cite{GECAM_FRB}, Solar Flares (SFLs) \cite{GECAM_SF}, as well as Terrestrial Gamma-ray Flashes (TGFs) and Terrestrial Electron Beams (TEBs) \cite{GECAM_TEB}.

GECAM-A/B were launched on December 10, 2020 (Beijing time) to a Low Earth Orbit (LEO) with an altitude of about 600\,km and inclination angle of 29 degrees. Two types of detectors are equipped on GECAM-A/B, namely gamma-ray detectors (GRDs) and charged particles detectors (CPDs) \cite{GEC_INS_An2022,GEC_INS_Xv2021}. Both GECAM-A and GECAM-B are equipped with 25 GRD detectors and 8 CPD detectors, nearly uniformly distributed on the hemispherical dome of the satellite. All detectors are pointed to different directions to achieve a very wide field of view. 

GECAM-C is one of the instruments onboard the SATech-01 satellite, which was launched on July 27, 2021. SATech-01 is operating in a Sun Synchronous Orbit (SSO) with an incidence angle of 97.4$^\circ$ and an altitude of about 500\,km. GECAM-C consists of two hemispherical domes; one sits on the top side of the satellite while the other is on the bottom side. Each dome is equipped with 6 GRDs and 1 CPD. For the top dome, the detectors are numbered from GRD1 to GRD6 and CPD1. For the bottom dome, the detectors are numbered from GRD7 to GRD12, along with CPD2. Due to the similar design of these two domes, each detector in the top dome of GECAM-C corresponds to a detector in the bottom dome that is oriented in the opposite direction (i.e. GRD1-GRD7, GRD2-GRD12, GRD3-GRD11, GRD4-GRD10, GRD5-GRD9, GRD6-GRD8 and CPD1-CPD2). Such a configuration of detectors provides a 4$\pi$ (full-sky) field of view for GECAM-C. 
In addition to GECAM-C, SATech-01 also carries many other payloads including the high-precision system for measuring the magnetic field based on a coherent population trapping magnetometer (referred to as the CPT system) \cite{SATech_01,CPT_system}.

GRDs are sensitive to photons including X-ray and $\gamma$-ray and also sensitive to charged particles, while CPDs are sensitive to charged particles but much less sensitive to photons. Therefore, considering the significance of signals on both the GRD and CPD, one can determine whether some signals are caused by charged particles or photons. All detectors on GECAM do not have a collimating structure, meaning that each detector has a FoV of about 2$\pi$. Moreover, there are no magnetic deflection or electric deflecting structures before the detectors for GECAM. 

Based on the assumption of parallel incidence of particles, which is always satisfied for astrophysical photon events such as GRBs, each GECAM can constrain the localization of high-energy transients independently. The basic idea is that for a certain incident direction, the incident angles of different detectors are different, resulting in modulated photon counts on different detectors by the incident angles. By comparing the counts from each detector, the incident direction can be inferred \cite{YIZ_LOC_GEC_2023}. 
In practice, the relative photon counts of each detector are calculated in advance and summarized as a template. Then we just need to compare the observation data with the template and choose the angle corresponding to the closest result. 

However this method for localization is not usually effective for events caused by charged particles. 
On one hand, charged particles do not always move in the same direction and they may simultaneously approach the satellite from all directions in many cases, especially for the particles precipitation events. 
On the other hand, since charged particles have circular motion in magnetic field and a non-zero pitch angle, the detected charged particles usually do not meet the assumption of parallel incidence.

In fact, for the detection of events caused by charged particles, the description of pitch angle is more appropriate than the incoming direction. 
Inspired by the localization method of photons, we noticed that different pitch angles will also result in different distributions of counts on different detectors. 
Thus we name this method as ``Pitch Angle based on Counts Distribution" (PACD). 

First, we normalized the counts of all the detectors by
\begin{equation}
    N_{det,norm}=\frac{N_{det}-N_{min}}{N_{max}-N_{min}}
\label{equ:norm_func}
\end{equation}
where $N_{det}$ is the raw net counts recorded by each detector, $N_{max}$ and $N_{min}$ are the maximum and minimum raw net counts of all the detectors respectively, and $N_{det,norm}$ is the normalized relative counts of each detector. 
The main purpose of this normalizing method is to highlight the relative differences in the signal strength between all the detectors as much as possible. 

The next step is to calculate the prediction of the relative counts of each detector by simulation. Since the assumption that the charged particles are ``local", which means the charged particles are present in all the space surrounding the satellite homogeneously, the needed input is only the direction of the local geomagnetic field. 
For GECAM-A/B, no instruments for detecting the geomagnetic field are equipped thus the International Geomagnetic Reference Field (IGRF) model \cite{IGRF_13} is used to calculate the parameters of the local geomagnetic field. 
For GECAM-C, since the CPT system is onboard the same satellite, the precise local geomagnetic field can be obtained from the direct measurement of the CPT system. 
By generating enough numbers of electrons based on different pitch angles, that are uniformly emitted from a spherical surface wrapping the satellite mass model, the expected counts of all detectors are simulated by the Geant4 toolkit. \cite{Geant4}, and are normalized by the same process in Equation,\ref{equ:norm_func}. The modified Shielding Physics List is utilized in the simulation, which has been successfully employed in the simulation of \textit{Insight}-HXMT \cite{HXMT_sim} and GECAM-D \cite{GTM_INS_wang2024}. 

The final step is to compare the simulation with the observation to get the pitch angle. The normalized relative counts of each detector can be treated as a vector. For GECAM-A/B, the vector has a dimension of 33 (25 GRDs and 8 CPDs) while the dimension is 14 (12 GRDs and 2 CPDs) for GECAM-C. The cosine similarity is used to calculate the consistency (goodness of fit) between the observation and the simulation, which is defined as
\begin{equation}
    {\rm cos}\theta=\frac{V_{sim} \cdot V_{obs}}{\lvert V_{sim}\rvert \lvert V_{obs}\rvert}
\label{equ:sim_func}
\end{equation}
where $V_{sim}$ and $V_{obs}$ are the vectors of simulation and observation respectively and cos$\theta$ is the cosine similarity.  
The pitch angle corresponding to the simulation with the cosine similarity closest to 1 is considered as the measurement result. 
Using cosine similarity is also to focus on the relative strength of signals between different detectors, trying to ignore the absolute amplitude of the signals.

Obviously, the pitch angle measurement resolution of PACD is affected by both the significance of the detected signal and the sampling interval of pitch angle in simulation, when only statistical errors are taken into consideration. 
However, a trade-off needs to be made between the calculation cost and the precision of the pitch angle: if the angle interval is reduced, the precision of the pitch angle will be improved, but the calculation cost will also increase rapidly.

\section{Observation and Analysis} \label{sec:obs}
\begin{figure}
\centering
\begin{overpic}[width=0.48\textwidth]{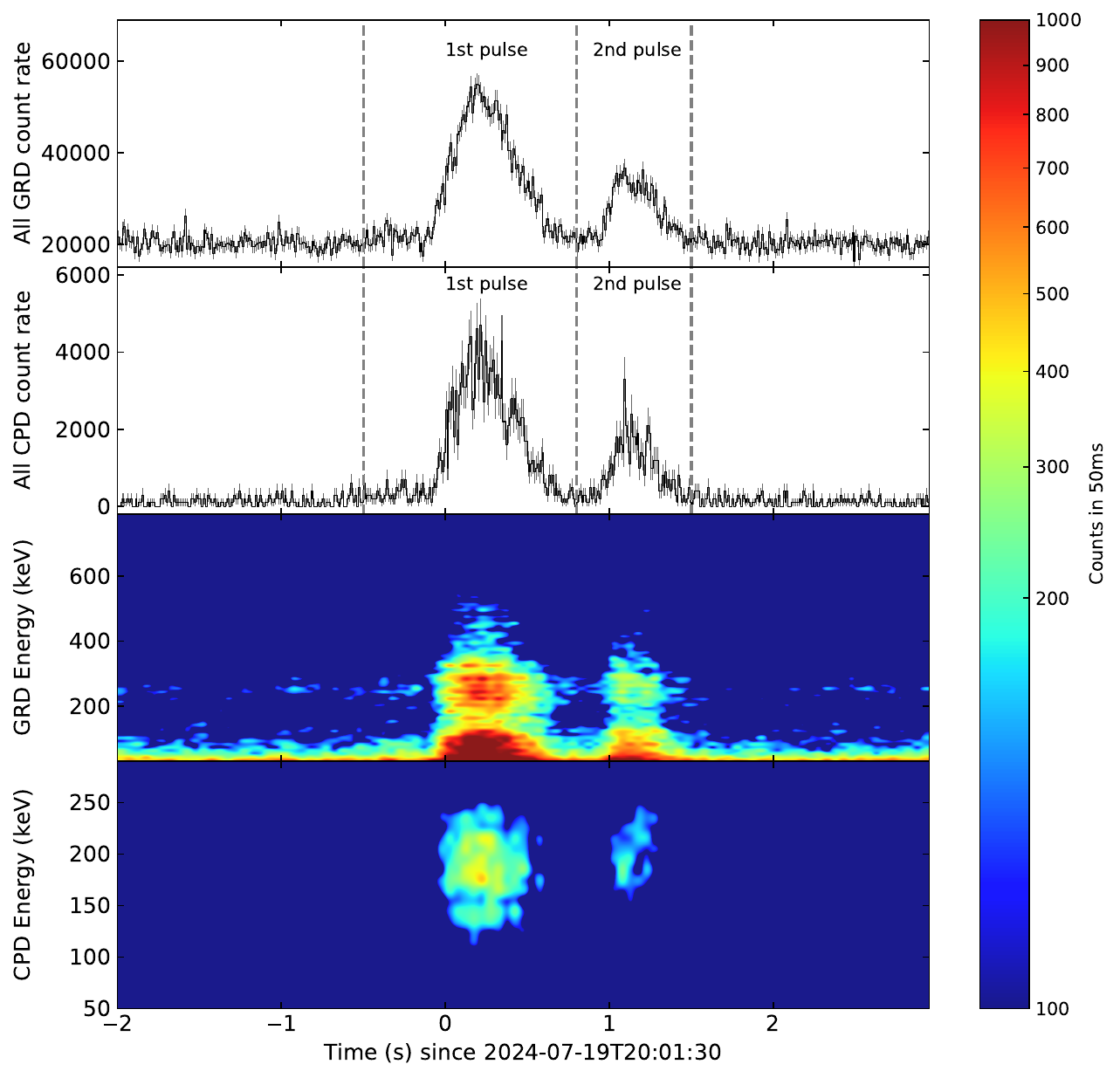}\end{overpic} 
\caption{\textbf{GECAM-C GRDs and CPDs lightcurve of tn240719\_200130.}}
\label{fig:Fig_total_lc}
\end{figure}

At 2024-07-19T20:01:30 (UTC, denoted as T$_0$), GECAM-C detected an interesting event, which is named tn240719\_200130 according to the GECAM convention of trigger event. With the event-by-event data of GECAM-C, the light curve shows that this event consists of two separated pulses, with a total duration of about 1.5\,s, as shown in Fig.\ref{fig:Fig_total_lc}. Considering the CPDs' lightcurve exhibits the same profile as GRDs' lightcurve, this signal is identified as electrons rather than gamma-ray photons or protons. 

\begin{figure}
\centering
\begin{tabular}{cc}
    \begin{overpic}[width=0.45\textwidth]{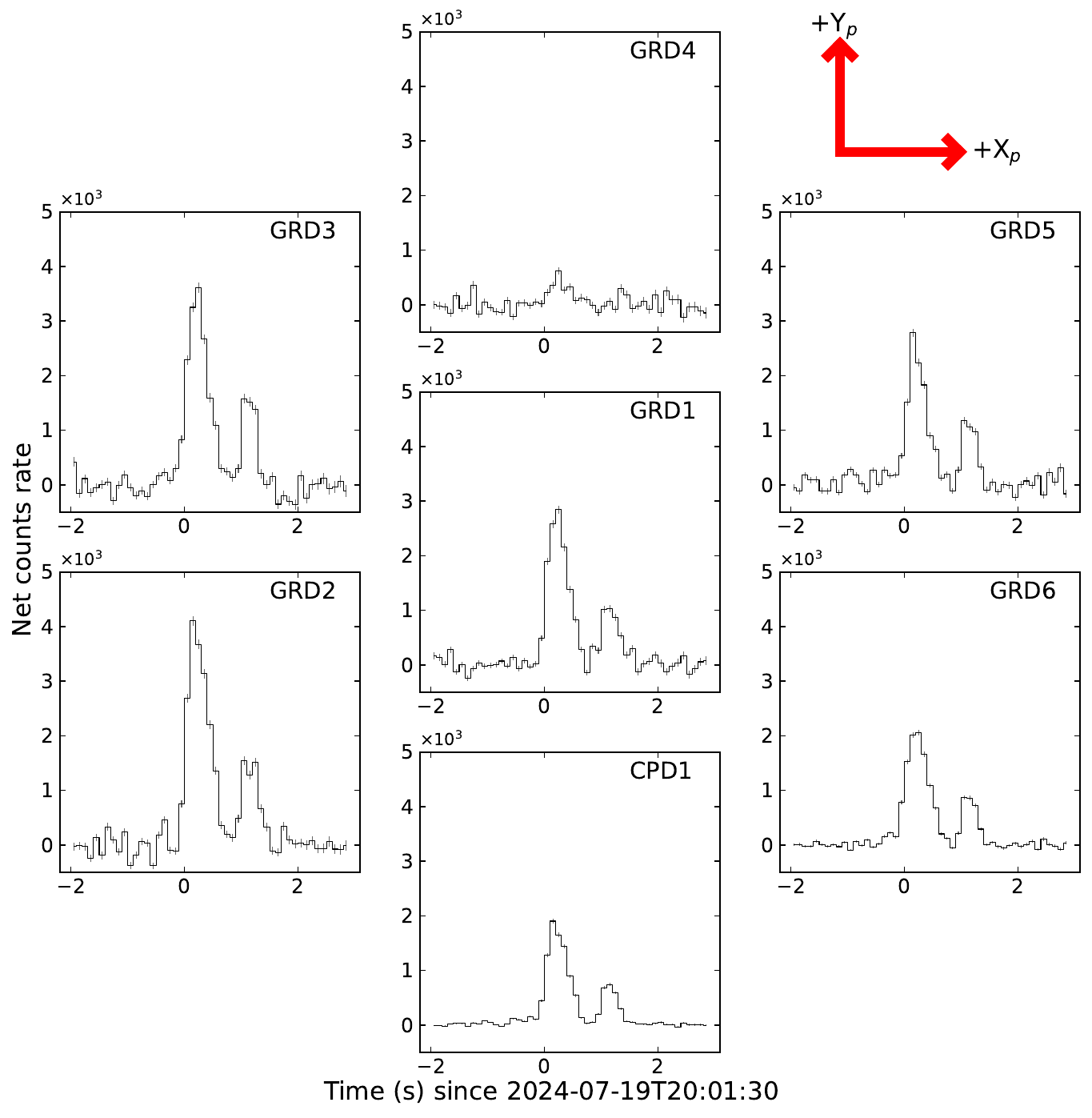}\put(-2, 90){\bf a}\end{overpic} &
    \begin{overpic}[width=0.45\textwidth]{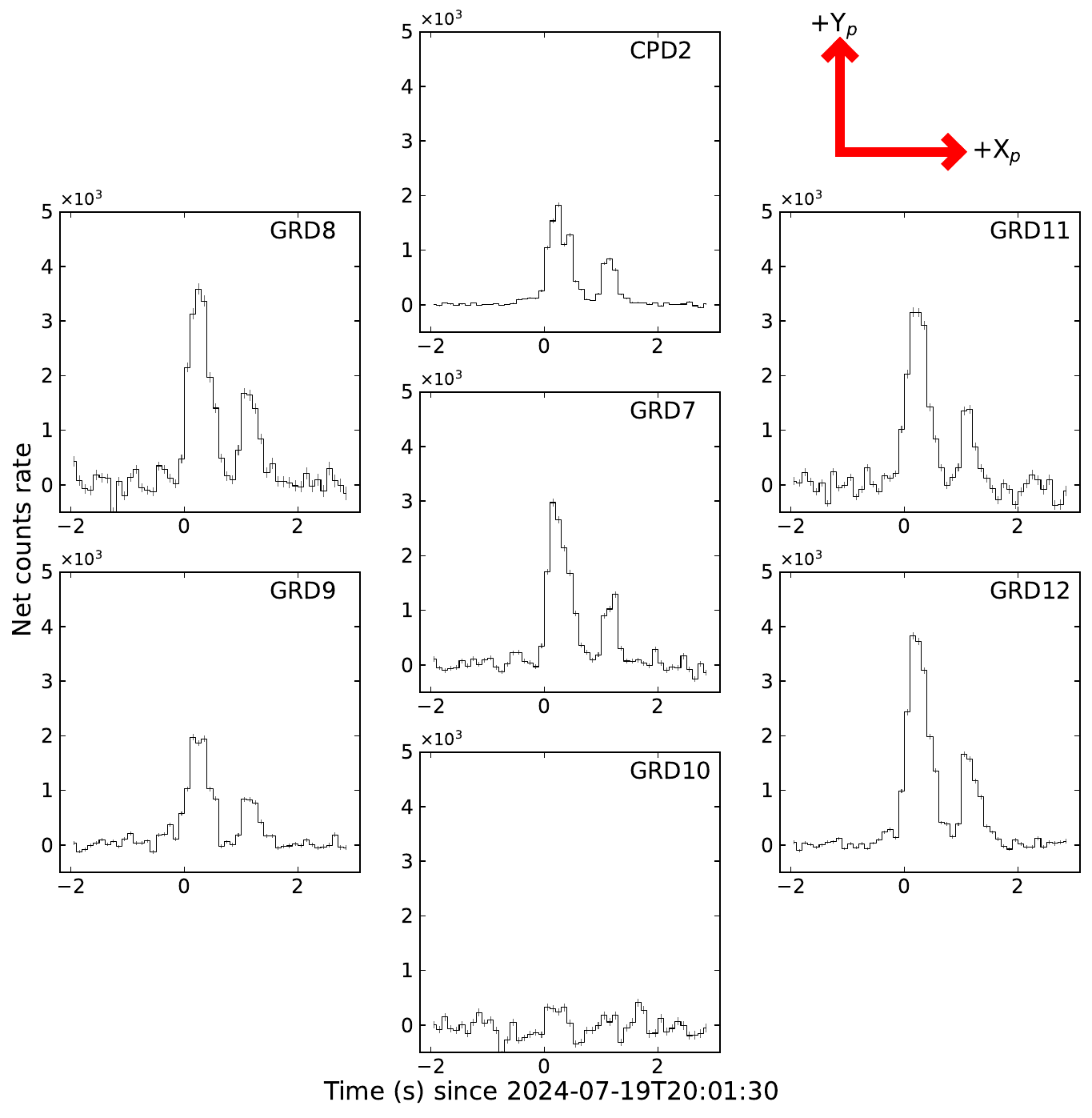}\put(-2, 90){\bf b}\end{overpic} \\
\end{tabular}
\caption{\textbf{Lightcurves of each detectors.} a, the top dome (+Z$\rm_p$). b, the bottom dome (-Z$\rm_p$). It is interesting that only two detectors, GRD4 and GRD10, with opposite orientation almost have no signal.}
\label{fig:Fig_dome_lc}
\end{figure}

Usually the duration of electron precipitation events is much longer than tn240719\_200130, while the duration of TEBs is about tens of milliseconds, much shorter than tn240719\_200130. 
Therefore, particle events with such a strange duration are relatively rare, which have caught our attention. 
However, the most interesting feature of tn240719\_200130 is that only two detectors, GRD4 and GRD10, with opposite orientation, almost have no signal, which can be seen in Fig.\,\ref{fig:Fig_dome_lc}. 
Considering the orientation distribution of the GECAM-C detectors, such a pattern of relative counts distribution can only occur when the velocity of the electron population is nearly parallel to the detector plane of GRD4 and GRD10, which is valid only when the pitch angle of the electron population is approximately 90$^\circ$. 
This provides compelling evidence of a unimodal, narrow distribution of pitch angle.
Therefore, this event is a good case to test our pitch angle measurement method.

For the pitch angle measurement, the first step is to obtain the local geomagnetic field parameters. 
The raw geomagnetic parameters in the Geodetic coordination system are converted to the payload coordination system of GECAM-C by the attitude during this event. 
The information of the magnet field, including strength and direction, is measured by the CPT system and listed in Table.\,\ref{tab:MG_information}.

\begin{figure}
\centering
\begin{overpic}[width=0.9\textwidth]{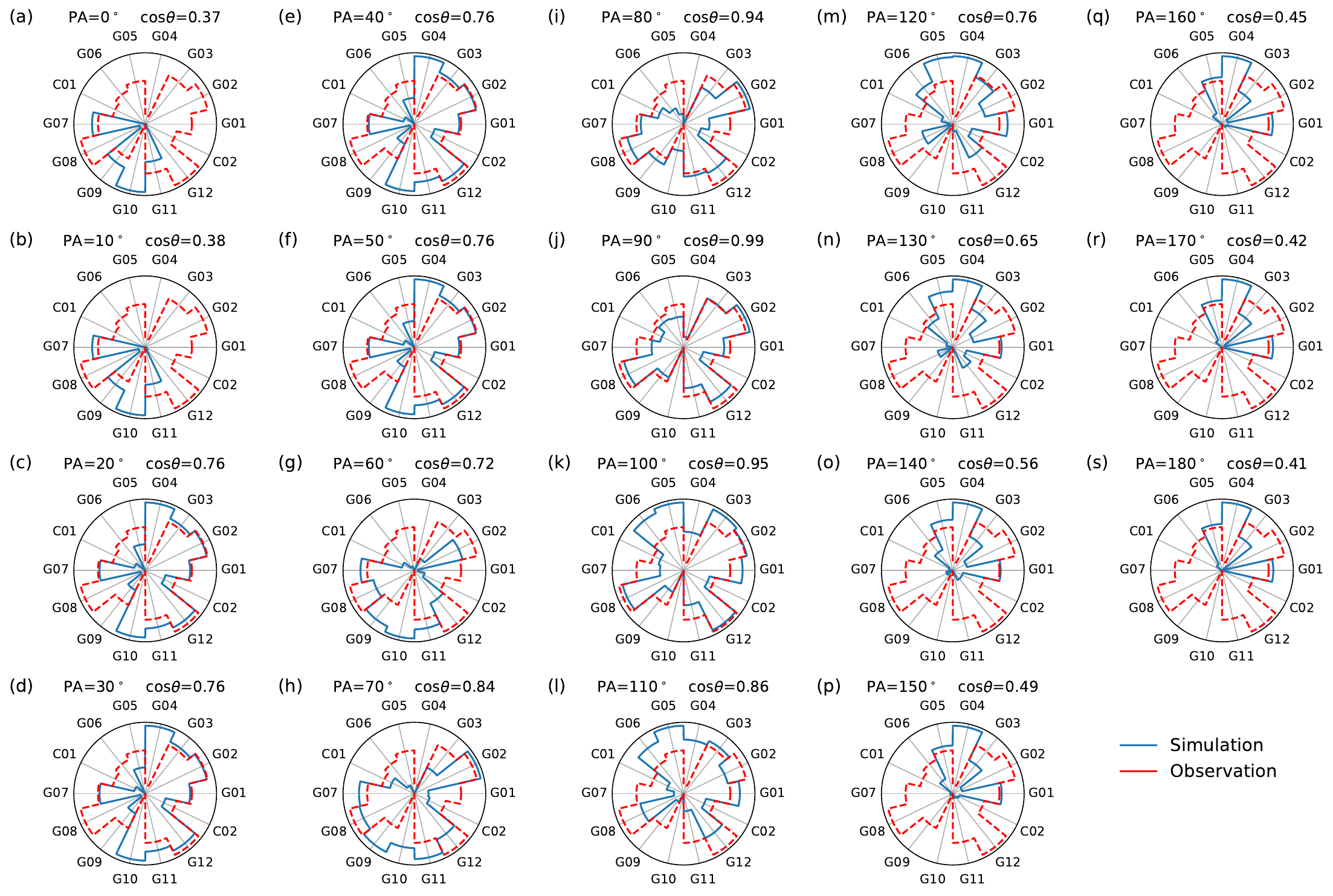}\end{overpic}
\caption{\textbf{Simulation compared with observation.} For each panel, the red dashed line represents the normalized relative counts for each detectors, while the blue solid line represents the normalized relative counts for each detectors at a fixed pitch angle. GRD and CPD are abbreviated as ``G" and ``C" respectively. ``PA" is the abbreviation of ``Pitch Angle". And cos$\theta$ is the cosine similarity.}
\label{fig:Fig_simulation}
\end{figure}

\begin{figure}
\centering
\begin{overpic}[width=0.9\textwidth]{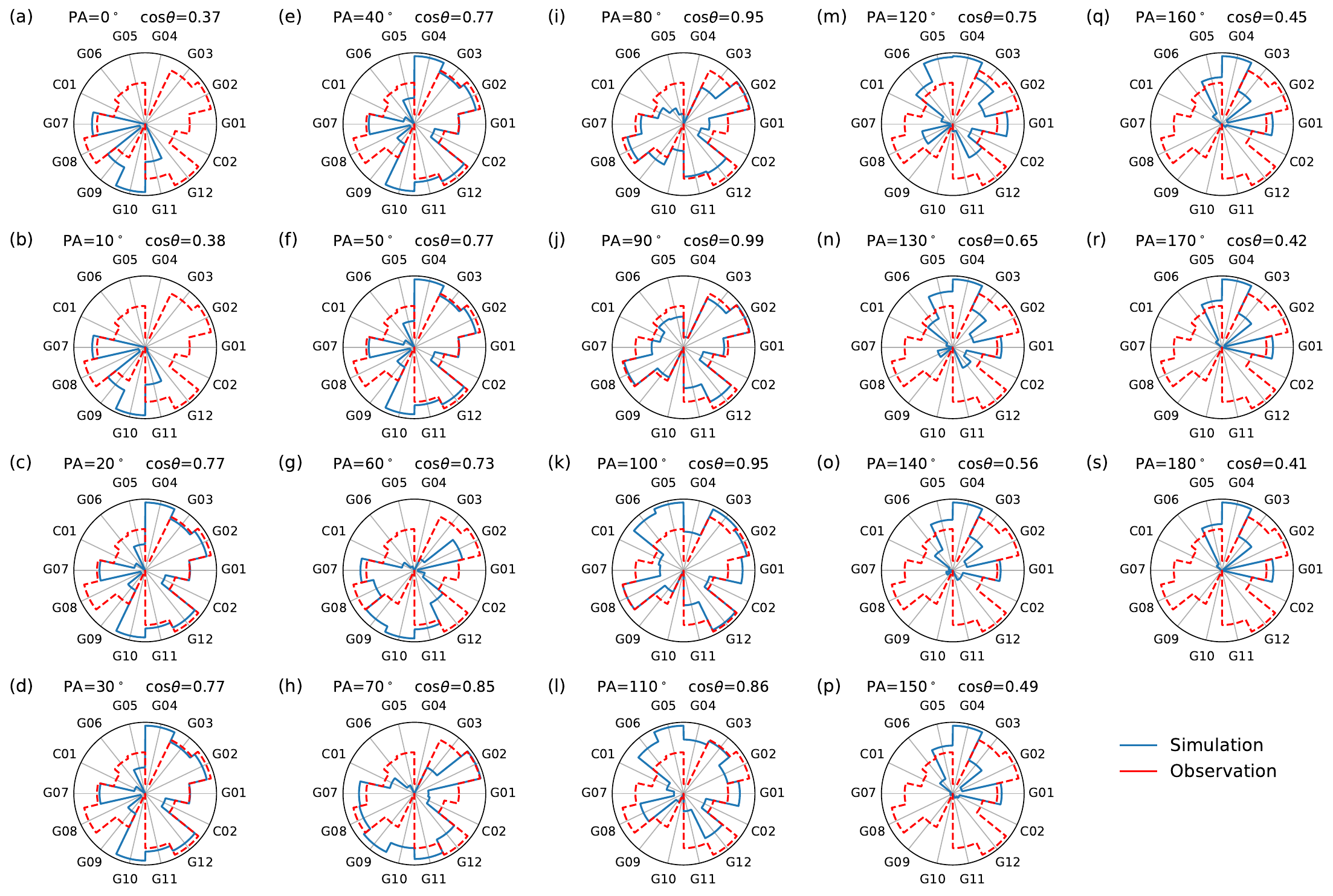}\end{overpic}
\caption{\textbf{Simulation compared with observation of the first pulse (T$_0$-0.5\,s $\sim$ T$_0$+0.8\,s) of tn240719\_200130.}}
\label{fig:Fig_simulation1}
\end{figure}

\begin{figure}
\centering
\begin{overpic}[width=0.9\textwidth]{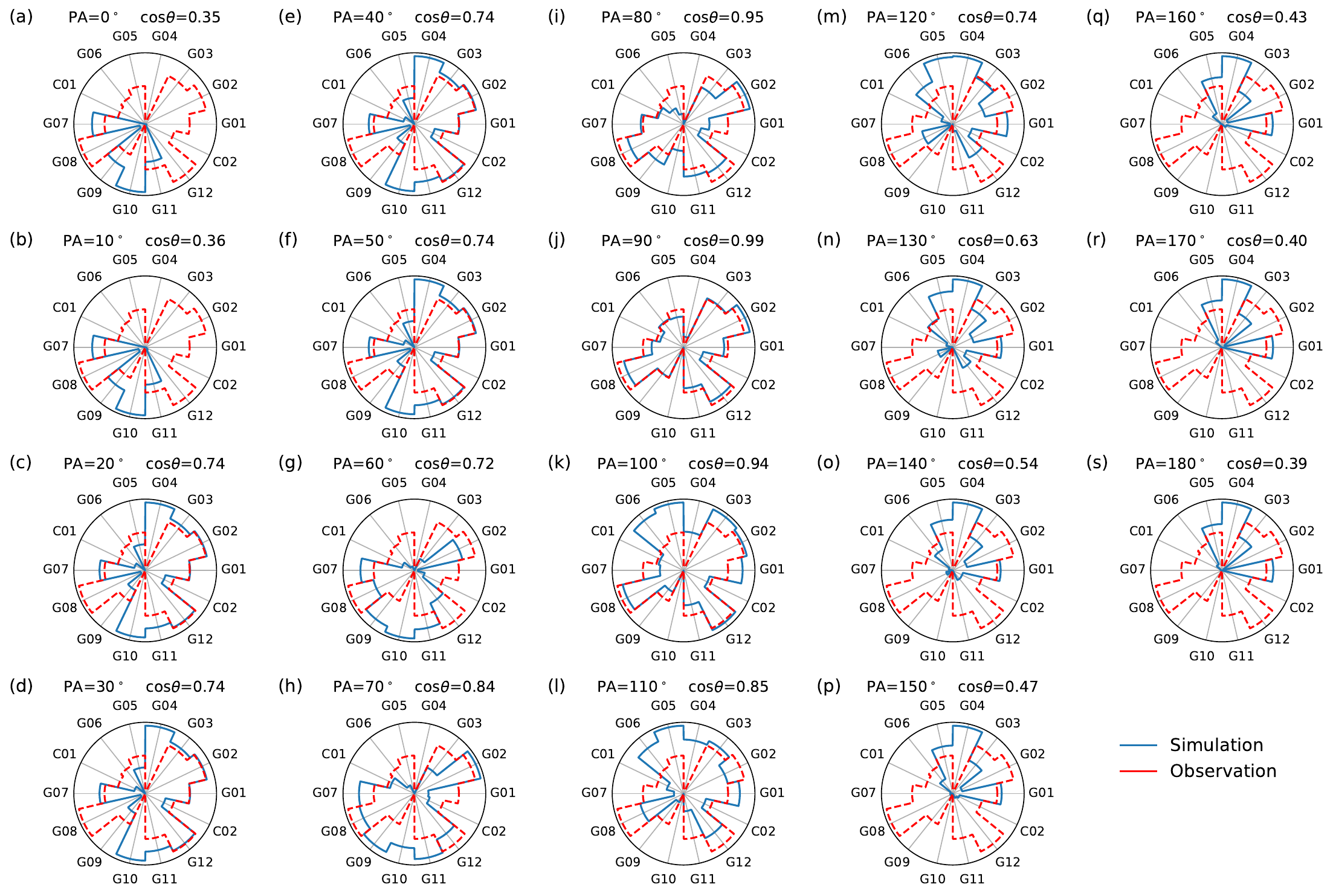}\end{overpic}
\caption{\textbf{Simulation compared with observation of the second pulse (T$_0$+0.8\,s $\sim$ T$_0$+1.5\,s) of tn240719\_200130.}}
\label{fig:Fig_simulation2}
\end{figure}

Since the best resolution of relativistic electron pitch angle observations is no better than 9$^\circ$ thus far \cite{pith_angle_overview}, we chose a 10$^\circ$ angle interval in the simulation. The energy of the incident electrons in the simulation is uniformly sampled from 1 to 2000 keV. 
Followed by the method described in Section.\ref{sec:ins}, the simulation results for different pitch angles are listed in Fig.\,\ref{fig:Fig_simulation}. The results indicate that the simulation of 90$^\circ$ pitch angle has the highest similarity with observation (Fig.\,\ref{fig:Fig_simulation}), and there are no other simulation results with the same high similarity, which is consistent with the prediction of unimodal pitch angle distribution of approximately 90$^\circ$. 
Hence the pitch angle of tn240719\_200130 is considered to be indeed 90$^\circ$. 
Additionally, we also investigated the pitch angle of the two pulses to explore whether there is any significant evolution. 
The results show that the pitch angle of both pulses is about 90$^\circ$ (Fig.\,\ref{fig:Fig_simulation1} and Fig.\,\ref{fig:Fig_simulation2}), with no significant differences, which is consistent with the lightcurve in Fig.\,\ref{fig:Fig_dome_lc}.

\begin{figure}
\centering
\begin{tabular}{cc}
    \begin{overpic}[width=0.45\textwidth]{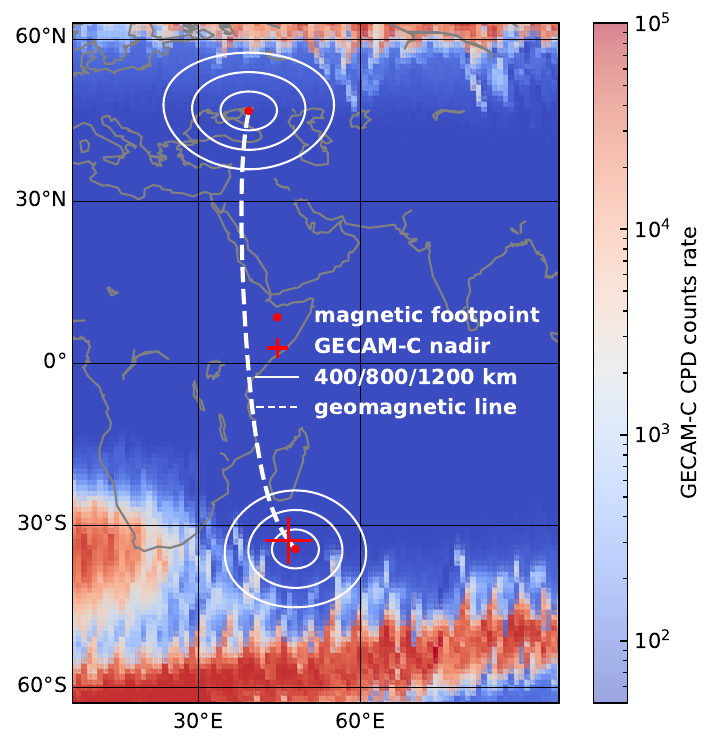}\put(-2, 90){\bf a}\end{overpic} &
    \begin{overpic}[width=0.45\textwidth]{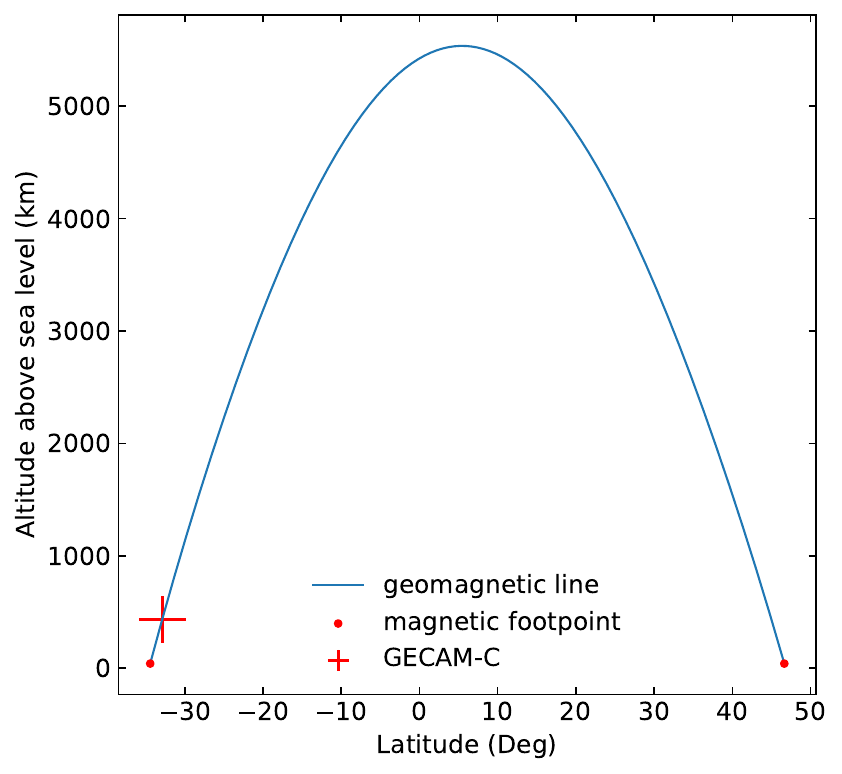}\put(-2, 95){\bf b}\end{overpic} \\
\end{tabular}
\caption{\textbf{Geomegnetic line tracing result.} a, the nadir of GECAM-C on the map during tn240719\_200130. b, the altitude of GECAM-C during tn240719\_200130. GECAM-C is very close to the southern magnetic footpoint.}
\label{fig:Fig_trace}
\end{figure}

Interestingly, a 90$^\circ$ pitch angle means that the electrons are about to be bounced. By tracing the geomagnetic field line with the International Geomagnetic Reference Field (IGRF-13) model \cite{IGRF_13}, we found that the south magnetic footprint is close to the nadir (Fig.\,\ref{fig:Fig_trace}). Thus a $\sim$90$^\circ$ pitch angle could be reasonable. But whether the second pulse is a bounced peak still cannot be confirmed, as the time interval between the two pulses is relatively longer than TEBs (usually about a few ms).

It also needs to be noted that the geomagnetic field parameters for tn240719\_200130 are from precise measurement by the CPT system. For some cases, there may be no accurate in-situ magnetic field measurement. For these cases, the geomagnetic parameters can be calculated from geomagnetic field model, such as IGRF model. 
The direction of geomagnetic field derived from the CPT system and IGRF model in GECAM payload coordinate system are both listed in Table.\,\ref{tab:MG_information}. And the difference between the two directions is not large, much smaller than the sampling interval of the pitch angle in the simulation. Therefore we conclude that the pitch angle measurement based on IGRF model is reliable.

\begin{table}[htbp]
\caption{\centering{Geomagnetic field parameters of tn240719\_200130 in Geodetic coordinate system}}
\resizebox{0.6\linewidth}{!}{
    \begin{tabular*}{\hsize}{@{}@{\extracolsep{\fill}}ccccccccccc|cc@{}}
    \cline{1-13}
      & Longitude & Latitude & Altitude & D & I & X & Y & H & Z & F & $\theta^a$ & $\phi^a$\\
    & (Deg) & (Deg) & (km) & (Deg east) & (Deg down) & (nT) & (nT) & (nT) & (nT) & (nT) & (Deg) & (Deg) \\
    \cline{1-13}
    CPT & 46.73 & -32.87 & 433.09 & -31.61 & -61.63 & 12009.90 & -7390.75 & 14101.80 & -26110.62 & 29675.22 & 41.11 & 83.31 \\
    IGRF-13 $^b$& 46.73 & -32.87 & 433.09 & -31.70 & -60.94 & 12256.38 & -7568.43 & 14404.86 & -25921.08 & 29654.72 & 41.77 & 83.58 \\
    \cline{1-13}
    \end{tabular*}
}
\\
$^a$The geomagnetic field direction payload coordinate system.\\
$^b$https://pypi.org/project/pyIGRF/
\label{tab:MG_information}
\end{table}

\begin{figure}
\centering
\begin{overpic}[width=0.48\textwidth]{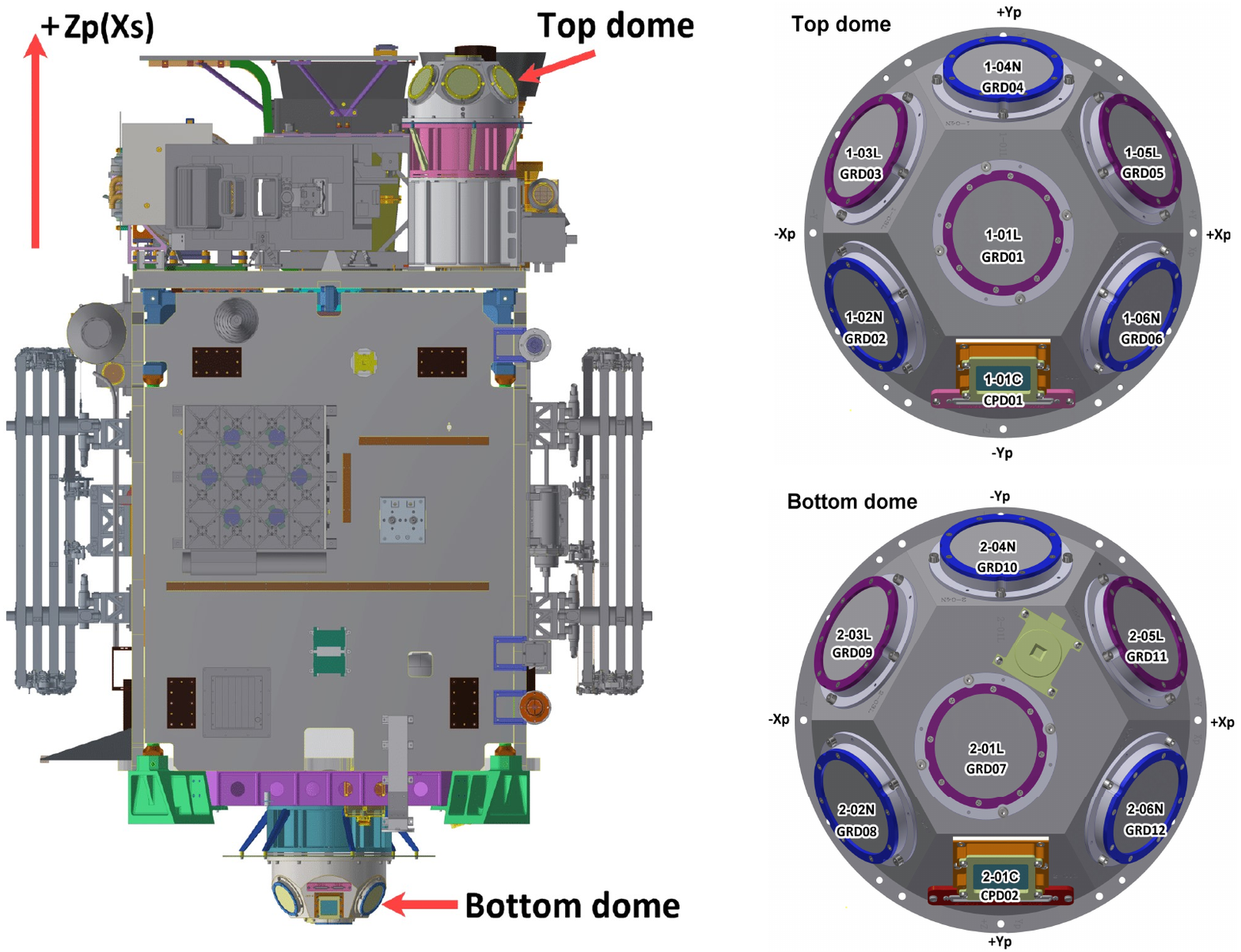}\end{overpic} 
\caption{Overview of SATech-01 satellite and two dome of GECAM-C. 
As for the legend of coordinate systems, the subscript ``p" represents the payload coordinate system, while the subscript ``s" represents the satellite coordinate system. The relation between the two coordinate system are: +X$\rm_p$=+Y$\rm_s$, +Y$\rm_p$=+Z$\rm_s$, +Z$\rm_p$=Z$\rm_s$. Due to the different orientations of the detectors, each detector in the top dome of GECAM-C corresponds to a detector in the bottom dome that is oriented in the opposite direction (i.e. GRD1-GRD7, GRD2-GRD12, GRD3-GRD11, GRD4-GRD10, GRD5-GRD9, GRD6-GRD8 and CPD1-CPD2).}
\label{fig:Fig_sat_total}
\end{figure}

\section{Discussion and Summary} \label{sec:sum}

For high energy transient monitors with large FOV like GECAM, although they are primarily designed to detect astronomical events, they have already played a key role in the detection of various charged particle events including TEBs and LEPs. 
However, for a long time, research on these charged particle events has faced certain limitations due to the design of the instruments. Especially, the information of the pitch angle is usually inaccessible from the observations of these all-sky monitors, however, the pitch angle is crucial in studying their generation and propagation of charged particles. 

In order to make full use of existing instruments and data for research, we developed a dedicated method for measuring pitch angle based on the characteristic of multi-detectors-with-different-orientations. 
As the first paper of this series, we mainly focus on the basic conception of PACD method and its feasibility, which has been shown by the analysis of a typical event observed by GECAM-C. 
We need to acknowledge that the basic version of the PACD method introduced in this paper has numerous limitations and is based on many idealized assumptions, including a unimodal pitch angle distribution, uniform energy distribution, and so on. However, these issues will be addressed in a subsequent series of articles, gradually enhancing the practicality and expanding the applicability of the PACD method.

One advantage of the PACD method is the variable and high time resolution, as shown in the pulse resolved pitch angle measurement in tn240719\_200130 (Fig.\,\ref{fig:Fig_simulation1} and Fig.\,\ref{fig:Fig_simulation2}). For specialized space physics instruments, especially those that measure the pitch angle by rotation, the time resolution is determined by the rotation period, which typically ranges from several seconds to even dozens of seconds \cite{pith_angle_overview}, far exceeding the duration of short timescale events like TEBs. Therefore, it is impossible to measure the temporal evolution of the pitch angle of TEBs by these instruments. While for the PACD method, the time resolution is mainly determined by the flux of charged particles, and the higher flux, the higher time resolution. Hence the PACD method is expected to measure the evolution of the pitch angle for bright TEBs and to test the generation and propagation model of TEBs.

The PACD method endows the ability to analyze the pitch angle of events caused by charged particles utilizing GECAM data. 
This will help study the generation and propagation mechanisms of phenomena such as TEBs and LEPs by measuring their pitch angle distributions. 
Moreover, though the pitch angle measurement result of an event detected by GECAM-C is presented as an example, PACD method is not only effective for GECAM, but also valid for other high-energy transient monitors with a wide field of view (such as \textit{Fermi}/GBM and SVOM/GRM).


\bmhead{Acknowledgments}
We acknowledge the support by 
the National Natural Science Foundation of China (Grant No. 12273042),
the Strategic Priority Research Program of the Chinese Academy of Sciences (Grant Nos. 
XDA30050000, 
XDB0550300 
),
the National Key R\&D Program of China (2021YFA0718500). 
The GECAM (Huairou-1) mission is supported by the Strategic Priority Research Program on Space Science (Grant No. XDA15360000) of Chinese Academy of Sciences. We appreciate Xuzhi Zhou, Xiaocheng Guo, Wenya Li, Xiaochao Yang, Qinghe Zhang, Gaopeng Lu for helpful discussions.

\subsection*{Funding}
The (co-)authors are funded by the funding the agencies described in the acknowledgment section. 

\subsection*{Conflicts of Interest}
The authors declare that the research was conducted in the absence of any commercial or financial relationships that could be construed as a potential conflict of interest.

\subsection*{Data availability statement}
The processed data are presented in the tables and figures of the paper, which are available upon reasonable request. The authors point out that some data used in the paper are publicly available.

\subsection*{Author's Contribution}
Chenwei Wang and Shaolin Xiong wrote the main manuscript text. 
Hongbo Xue, Yiteng Zhang and Shanzhi Ye processed the data of CPT system. 
All authors participated in the discussion and reviewed the manuscript.

\clearpage
\bibliography{sn-bibliography}

\end{document}